\documentstyle[sprocl,epsf]{article}

\def\Journal#1#2#3#4{{#1} {\bf #2}, #3 (#4)}

\begin{document}

\title{   TESTING COSMOLOGICAL VARIATIONS 
       \\ OF FUNDAMENTAL PHYSICAL CONSTANTS 
       \\ BY ANALYSIS OF QUASAR SPECTRA}

\author{ D.A.~VARSHALOVICH, A.Y.~POTEKHIN, A.V.~IVANCHIK }

\address{Ioffe Physical-Technical Institute
                    \\ St.-Petersburg 194021, Russia}

\author{ V.E.~PANCHUK }

\address{6-m Telescope Observatory (SAO), 357147, Russia}

\author{ K.M.~LANZETTA }

\address{State University of New York at Stony Brook, NY 11794--2100, 
USA}

\maketitle\abstracts{
Contemporary multidimensional cosmological theories predict 
different variations
of fundamental physical constants in course of the cosmological
evolution. On the basis of the QSO spectra 
analysis, we show 
that the fine-structure
constant $\alpha = e^2 / \hbar c$ and the proton-to-electron
mass ratio $\mu = m_p / m_e$ reveal no statistically significant 
variation over the last $90\%$ of the lifetime of the Universe. 
At the $2\sigma$ significance level, the following upper bounds 
are obtained for the epoch corresponding to the cosmological 
redshifts $z\sim 3$ (i.e., $\sim 10$ Gyr ago):
$
 |\Delta\alpha / \alpha| < 1.6\times10^{-4} \quad\mbox{and}\quad 
 |\Delta \mu / \mu| < 2.2\times10^{-4}. 
$
The corresponding upper limits to the time-average rates of the 
constant variations are 
$$
 |\dot \alpha / \alpha| < 1.6\times10^{-14}\mbox{~yr}^{-1} 
\qquad\mbox{and}\qquad
 |\dot \mu / \mu| < 2.2\times10^{-14}\mbox{~yr}^{-1}.
$$ 
These limits serve as criteria for selection of those
theoretical models which predict $\alpha$ and $\mu$ variation
with the cosmological time. 
In addition, we test a possible anisotropy of the high-redshift 
fine splitting over the celestial sphere, which might 
reveal a non-equality of $\alpha$ values in causally disconnected 
areas of the Universe. 
}

\section{Introduction}
Contemporary theories (SUSY GUT, superstring and others) 
not only predict the dependence of fundamental physical 
constants on energy, but also 
have cosmological solutions in which low-energy values of 
these constants vary with the cosmological time. 
The predicted variation at the present epoch is small but non-zero, 
and it depends on theoretical model 
(see, e.g., Ref.~[\,\,\cite{vp95}\,] for references). 
In particular, 
Damour and Polyakov \cite{dp94} have developed
a modern version of the string theory, whose parameters 
could be determined from 
cosmological variations of the coupling constants and
hadron-to-electron mass ratios. 
Therefore observational 
tests of variability of fundamental constants may serve 
as an important tool for selection of the theoretical models. 

Quasar spectra are an important source of our knowledge of physical
conditions at early cosmological epochs, related to the redshifts up to 
$z \sim 4$. In particular, values of physical constants can be
extracted from quasar spectroscopic data and compared with the laboratory
values at the present epoch. This 
extragalactic information significantly supplements 
results obtained within the Solar system, 
which cover much smaller range of the redshifts, $z<0.2$. 
Although astrophysical measurements are generally less accurate than 
laboratory ones, the large cosmological time scales allow to obtain 
more stringent estimates of the variation rates of physical  
constants. 
%%%%%%%%%%%%%%%%%%%%%%%%%%%%%%%%%%%                             SECTION
\section{Proton-to-Electron Mass Ratio}
Electronic, vibrational, and rotational energies of the H$_2$ molecule
each display a different dependence on the reduced mass $m_p / 2$.
Therefore comparison of the wavelengths of various
electronic-vibrational-rotational molecular lines observed
in the spectrum of a high-redshift quasar with the corresponding molecular
lines observed in laboratory may reveal or limit variation
of $\mu = m_p / m_e$. One may obtain a quantitative
limit on such variation, provided that the sensitivity coefficients $K_i$ of
wavelengths $\lambda_i$ with respect to the $\mu$ variation are
known for each of these lines \cite{vp95}. 
If the value of $\mu$  at the early epoch $z$ 
of the QSO absorption spectrum
formation were different from the contemporary one, then the wavelength
ratios would deviate from unity
\begin{equation}
\frac{(\lambda_i / \lambda_k)_z}{(\lambda_i / \lambda_k)_0} \simeq
1+(K_i-K_k)\left(\frac{\Delta\mu}{\mu}\right)
\end{equation}
The object suitable for the analysis is the H$_2$ absorption system
toward PKS 0528-250 at the redshift $z=2.811$. Recently, 
Cowie and Songaila \cite{cs95} observed 
this quasar with Keck Telescope and arrived at
the 95\% confidence interval $-5.5\times10^{-4}<\Delta \mu / \mu<
7\times10^{-4}$, based on an analysis of 18 radiative transitions for H$_2$ 
molecule. 

We have performed a $\chi^2$ profile fitting analysis of a high-resolution
spectrum of PKS 0528-250, obtained with the Cerro-Tololo
Inter-American Observatory (CTIO) 4-meter telescope \cite{cl96}.
We have calculated the wavelength-to-mass sensitivity coefficients for a
larger number of spectral lines and employed them in the analysis. 
A total of 59 transitions for H$_2$ are incorporated into the $\chi^2$ fit,
and the absorption lines corresponding to these transitions occur across
the linear, saturated, and damped parts of the curve of growth. The redshift,
Doppler parameter, and column densities of the H$_2$ rotational levels
were adopted as free parameters.

A limit to variation of the proton-to-electron mass ratio was
sought by repeating the $\chi^2$ profile fitting analysis with an additional
free parameter $\Delta \mu / \mu$. The resulting parameter estimate and
$1 \sigma$ uncertainty is
\begin{equation}
 \Delta \mu / \mu = \left (8.3^{+6.6}_{-5.0} \right) \times10^{-5}.
\label{mu}
\end{equation}
This result indicates a value of $\Delta \mu / \mu$  that differ from zero
at the $1.6 \sigma$ level. The $2 \sigma$ confidence interval to
$\Delta \mu / \mu$ is $(-0.2,+2.2)\times10^{-4}$. 

\begin{figure}[t]
\begin{center}
\epsfysize=35mm 
\epsfbox{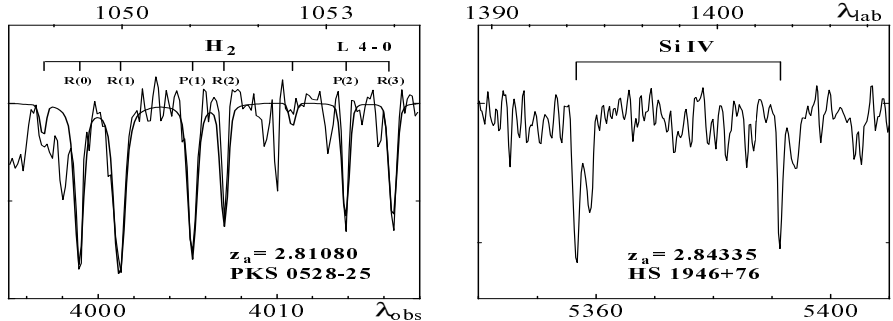}
\end{center}
\caption[]{Fragments of QSO spectra 
obtained with CTIO (left) and SAO (right) telescopes 
and used for estimations
of $\mu$ and $\alpha$ at the epochs $z_a$.}
\end{figure}
%%%%%%%%%%%%%%%%%%%%%%%%%%%%%%%%%%%                             SECTION
\section{Fine-Structure Constant}
Spectral observations of fine-splitted lines in quasar spectra 
provide the most direct test for variations of $\alpha$. 
 The relative splitting 
$(\Delta \lambda / \lambda)_z$ is proportional to $\alpha^2$ 
at the redshift $z$.
Therefore, the ratio 
\begin{equation}
\frac{(\Delta \lambda / \lambda)_z}{(\Delta \lambda / \lambda)_0} \simeq
1+2\left(\frac{\Delta \alpha}{\alpha}\right)
\end{equation}
would deviate from unity if the $\alpha$ value 
had changed.

In previous work \cite{pv94}, we 
have composed a catalogue of $\sim 1500$ 
alkali-like doublet wavelengths with 
$z>0.2$ observed in the quasar absorption spectra, and obtained 
the restriction 
$
|\Delta\alpha/\alpha| < 1.5\times 10^{-3}$ at $z \sim 2.5$.
In addition to the $z$-dependence of $\alpha$ averaged over the celestial 
sphere, we have also checked a possible spatial anisotropy of 
fine-splitting values at large $z$ \cite{vp94,vp95}. 
Within a relative statistical error $3\sigma < 0.3$\% the values of 
$\alpha$ turned out to be the same in different celestial 
hemispheres, which corresponds to their equality in causally disconnected areas. 
However, at the $2\sigma$ level a tentative anisotropy of estimated 
fine-splitting values has been found.  

In order to obtain a more stringent estimate of $\Delta\alpha / \alpha$,
we have carried out new spectral observations of quasars with 
the Special Astrophysical Observatory (SAO) 6-meter telescope  
\cite{vpi96}. 
We have observed several absorption systems in spectra of  
HS 1946+76, S5 0014+81, and S4 0636+68. 
Fine-structure 
doublet wavelengths of Si\,IV were measured 
with high resolution and accurate calibration,
using the same equipment and the same reduction procedures 
in order to obtain homogeneous data. 
The overall estimate at $z=2.8-3.1$ reads
\begin{equation}
(\Delta\alpha/\alpha)_z = (2\pm7)\times10^{-5},
\label{alpha}
\end{equation}
suggesting no statistically significant variation of $\alpha$ 
and providing the most stringent constraint to it. 
%%%%%%%%%%%%%%%%%%%%%%%%%%%%%%%%%%%                             SECTION
\section{Summary }
Our analysis of QSO spectra 
reveals no statistically significant variation 
of the fine-structure constant $\alpha=e^2/\hbar c$ 
and the proton-to-electron mass ratio
$\mu=m_p/m_e$. At the $2\sigma $ ($95\%$-significance) level, the 
estimates (\ref{mu}) and (\ref{alpha}) are obtained 
for the epoch corresponding to the cosmological redshifts $z \simeq 3.0$. 
These constraints are stronger than other recent  
astronomical \cite{cs95} and laboratory \cite{prestage95} 
results,
and they serve as effective criteria for selection of GUT 
and superstring models.  In particular, they enable to 
exclude the Teller--Dyson hypothesis \cite{dyson72} 
of logarighmic time-dependence of the fine-structure constant. 

\section*{Acknowledgments}
This work has been partly supported by RBRF (grant 96-02-16849a)
and by the Research Center ``Cosmion''.

\end{document}